\documentclass{caosp309}

\usepackage{graphicx}

\usepackage{natbib}
\articleNo{301}
\pubyear{2020}
\volume{50}
\volnumber{3}
\firstpage{1}
\received{May 1, 2020}
\accepted{July 28, 2020}

\def\BibTeX{{\rm B\kern-.05em{\sc i\kern-.025em b}\kern-.08em
             T\kern-.1667em\lower.7ex\hbox{E}\kern-.125emX}}

\begin{document}

%
\hauthor{A.\,Moharana at al.}

\title{Formation of compact hierarchical triples}


%
%
\author{
        A.\,Moharana\inst{1,2}\orcid{0000-0002-9616-512X}
      \and
        K.G.\,He{\l}miniak\inst{1}\orcid{0000-0002-7650-3603}
      \and 
        T.\,Pawar\inst{1}
       \and
       G.\,Pawar \inst{1}\orcid{0000-0003-3639-9052}
       }

%
\institute{
           Nicolaus Copernicus Astronomical Center, Polish Academy of Sciences, ul.                   Rabia\'{n}ska 8, 87-100 Toru\'{n}, Poland 
         \and 
          Astrophysics Group, Keele University, Staffordshire, ST5 5BG, UK\\
           U.K. 
          }

\date{March 8, 2003}

\maketitle

\begin{abstract}
Compact hierarchical triples (CHTs) are triple stars where the tertiary is in an orbit of period less than 1000 d. They were thought to be rare but we are discovering more of these systems recently, thanks to space-based missions like TESS, Kepler, and GAIA. In this work, we use orbital parameters obtained from these missions to constrain the formation process of CHTs. We also use spectroscopic and systemic parameters from our work, and the literature to understand the effects of metallicity and dynamics on the formation processes. 
\end{abstract}

%
\section{Introduction}
\label{intr}
Around 8\% of all Solar-type stars exist in triple systems \citep{SolarMult_raghavan}. This percentage increases are we go towards high-mass stars. This has increased interest in the detection, evolution, and formation of triple stars. New observations from space observatories have led to an exponential increase in the detection of these systems. This also led to the discovery of a new class of these triples, called compact hierarchical triples (CHTs; see \citealt{CHTrev_Borkovits} for a review). CHTs are triple stars where the tertiary star is in an orbit of period less than 1000 d. This roughly corresponds to an orbit size of less than 5 AU. This compactness causes dynamical changes which can be monitored over a period of a few years. Therefore, the effect of dynamics in the evolution of these systems can be constrained \citep{Eisner_CHT}.

The sizes of CHTs are on the scales of planetary systems. Therefore, they are interesting sites for studying star formation. The current idea of triple star formation is that they form through the process of sequential disk-instability (DI+DI), where the tertiary is formed from disk instabilities in the circumbinary disk around the inner binary \citep{OrbitsTriples_Tokovinin}. This was based on the observations of orbit alignment in triple stars. However, the analysis was done on triples whose semi-major axes were over 10 AU. In this work, we use estimates of orbital parameters of the present sample of CHTs to study the signatures of processes that are involved in the formation and evolution of CHTs. 

\section{Measurement of CHT parameters}
\begin{figure}
\centerline{\includegraphics[width=0.75\textwidth,clip=]{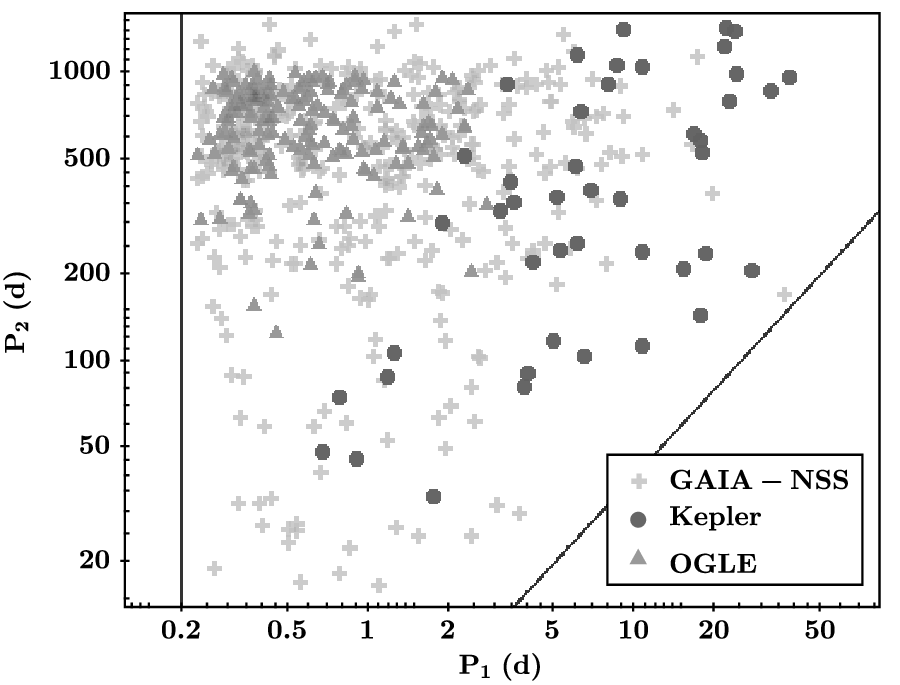}}
\caption{Distribution of orbital periods for the sample of CHTs used in this work. The grey lines mark the stability limits. The horizontal line on the left part of the plot represents the period limit for contact binaries while the slanted line on the right represents the limit for dynamical stability.}
\label{fig:pdist}
\end{figure}

Detecting stellar companions using EBs also allows us to measure the basic properties of the companion. While in some cases, we can only extract the idea about its orbit and lower mass limits, in some cases we can get all the stellar parameters of the tertiary. 
The first large sample of CHTs was reported by \cite{Kepler_Borko}, where the authors used eclipse timing variations (ETVs) from Kepler EBs\footnote{\url{https://keplerebs.villanova.edu/}} to characterise the orbits of 222 triple systems. Out of 222 of these triples, 110 are CHTs. However, only 45 CHTs have robust solutions with observations which cover more than two times the outer orbital period. Following the same methodology, \cite{OGLE1_Hajdu} gave us 258 CHTs in the Galactic Bulge, using EBs from OGLE \citep{OGLEEB_Sozynski} from which we selected 177 CHTs with robust solutions. The parameters that these studies estimated are the EB orbital period ($P_1$), tertiary orbital period ($P_2$), projection of the outer semi-major axis ($a_{2}\sin{i_2}$), tertiary orbital eccentricity ($e_2$), argument of periastron of the tertiary orbit ($\omega_2$), and the tertiary mass function ($f(m_3)$). 
\cite{GAIA_Czava} used a different methodology to estimate around 376 CHTs. The Gaia collaboration released a catalogue of DR3 non-single-star orbital solutions (GAIA-NSS; \citealt{GAIA_NSS}). \cite{GAIA_Czava} crossmatched the targets with EB catalogues from different sky surveys. They found that some systems have periods substantially longer than the EB orbital periods. The GAIA-NSS in most of these cases are likely to belong to outer orbits of tertiary stars in those systems. The parameters resulting from this survey were $P_1$, $P_2$, $a_{2}\sin{i_2}$, $e_2$, and $\omega_2$. The $P_1$ and $P_2$ distribution of the sample is presented in Fig.\ref{fig:pdist}.

\section{Calculation of mass ratios}
To get an estimate of the mass ratio, we need to connect the orbital parameters with masses using the tertiary mass function derived from Kepler's law,
\begin{equation}
\label{eq:massf}
    f(m_3) = \frac{m_3^3\sin^3{i}}{(m_\mathrm{bin}+m_3)^2} = \frac{P_2K^3}{2\pi G}(1-e_{2}^2)^{3/2}
\end{equation}

where $m_\mathrm{bin}$ is the total binary mass, and $K$ is the semi-amplitude of RV signal or ETV signal.
Eq.\ref{eq:massf} can be transformed into a cubic equation of $q_3=m_C/m_\mathrm{bin}$ as the variable with the coefficients being functions of $f$ and   $m_\mathrm{bin}$. For the ETV surveys, we have estimates of $f$. Therefore in such a case, it is easy to numerically solve the following cubic equation, assuming a  $m_\mathrm{bin}$,
\begin{equation}
\label{eq:massf3}
   m_\mathrm{bin} q_3^3 - fq_3^2 - 2fq_3 -f = 0
\end{equation}

Solving this equation for parameters estimated from \cite{GAIA_Czava} is not straightforward as $f$ is not available. The only unknown in the expression for $f$ is $K$. 
For our approximation, we set 
\begin{equation}
    K = 2\pi a_2 / P_2
\end{equation}
 where $a_2$ is available from GAIA-NSS solutions. From Eq.\ref{eq:massf} and taking $e_2=0$ for a simpler calculation, we get,

\begin{equation}
\label{eq:massfgaia}
    f_\mathrm{GAIA}(m_3)  = \frac{4\pi^2 a_2^{3}}{G P_2^{2}}
\end{equation}
 
We then estimate``ad-hoc" $q_3$ distribution from all the surveys, assuming $m_\mathrm{bin} = 2 M_{\odot}$. This assumption is based on the approximate idea that most of the known $m_\mathrm{bin}$ for the sample is spread around 2 $M_{\odot}$. The distribution was created by dividing the $q_3$ distribution into bins of size of 0.05. We also created an eccentricity distribution with a bin size of 0.03. 

\section{Discussion and conclusions}
We obtain distributions of eccentricity and tertiary mass ratios of all the CHTs from the large sample surveys, GAIA-NSS, OGLE, and Kepler. we also estimate the individual distributions of each of the surveys. The similarities and dissimilarities of the distributions are discussed in the successive sections.  
\subsection{Eccentricity distribution of the tertiary orbit}
\begin{figure}
\centerline{\includegraphics[width=0.7\textwidth,clip=]{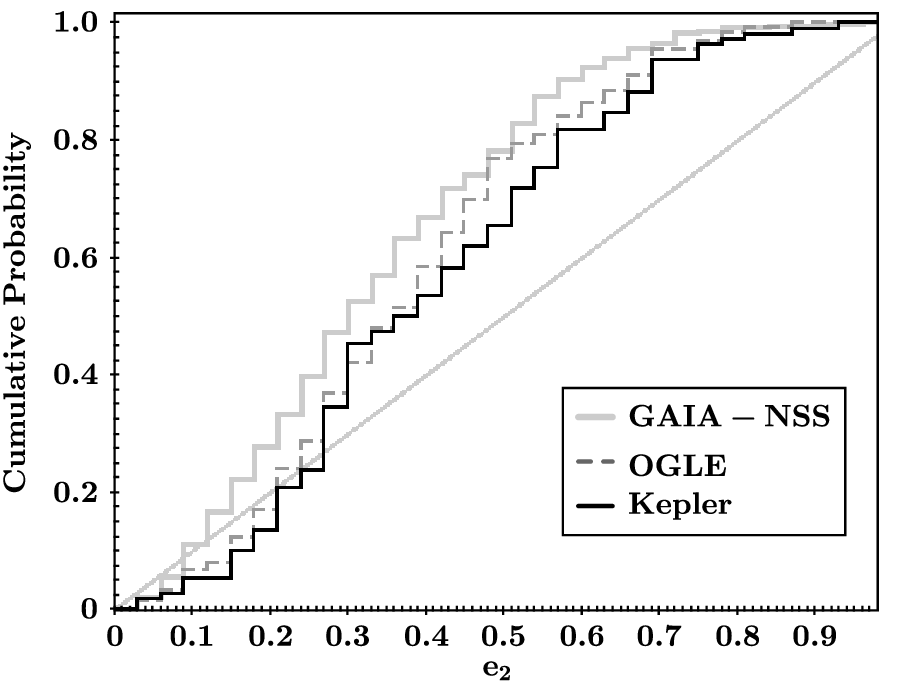}}
\caption{Cumulative probability distribution of $e_2$ for the sample of CHTs. The sub-samples of GAIA-NSS, OGLE, and Kepler are represented by grey-solid, grey-dashed, and black-solid lines respectively. The slanted grey line represents the expected trend for a flat distribution.}
\label{fig:eccdist}
\end{figure}

The eccentricity distribution has a specific shape which shows cumulative probabilities of $e_2$ larger than that expected from a flat distribution (Fig.\ref{fig:eccdist}). This is consistent with what was reported in \cite{GAIA_Czava} and was also observed in the eccentricity distribution of close binaries \citep{SolarMult_raghavan}. But if we look at the $e_2$ distribution closely, we see a small variation of trends between the different surveys. Kepler and OGLE distributions tend to go sub-flat at eccentricities 0.1-0.25. Such a trend was observed for metal-poor CHTs in \cite{Moharana_ST3n2}. Therefore this points at the Kepler sample being metal-poor. But OGLE also has a distribution similar to  GAIA at eccentricities around 0.45-0.55. This could be a sign that OGLE has CHTs with distinct metallicities and not a homogenous metallicity distribution like GAIA. 

\subsection{Bi-modality in mass ratio distribution}
\begin{figure}
\centerline{\includegraphics[width=\textwidth,clip=]{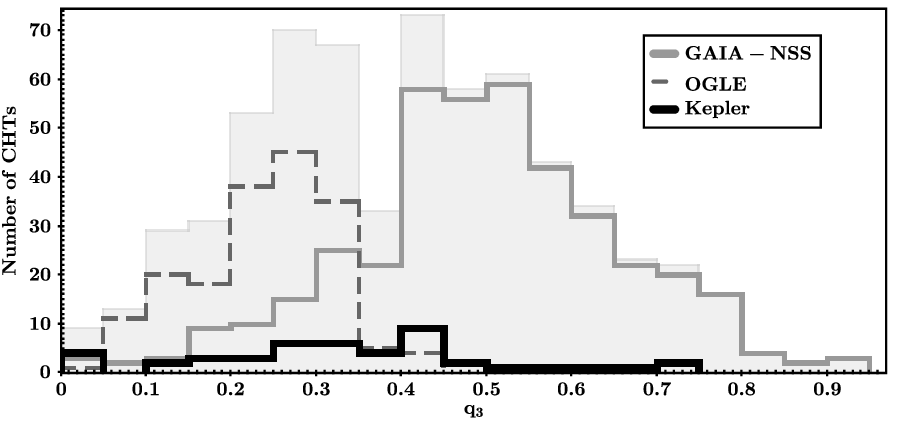}}
\caption{Distribution of $q_3$ for the large sample of CHTs (shaded histogram). The sub-samples of GAIA-NSS, OGLE, and Kepler are represented by grey-solid, grey-dashed, and black-solid lines respectively.}
\label{fig:mdist}
\end{figure}

One of the expected outcomes of the DI+DI scenario of star formation is that it produces ``twins", meaning that the mass ratio of the outer hierarchy to the inner hierarchy is 1. While deviations from 1 are expected, in our distribution we see bimodality, with two peaks at $q_3$ values smaller than 1 (Fig.\ref{fig:mdist}). This bi-modality was seen in \cite{Moharana_ST3n2} where we found that young CHTs have $q_3$ closer to 1 while old CHTs have low $q_3$. This hints that the two peaks in our distribution are due to two populations of different ages. The lower peak between 0.2 and 0.35 is mostly due to the OGLE sample. Therefore, the OGLE sample, which consists of CHTs in the Galactic Bulge, are old CHTs with their age in the order of giga-years. 

\subsection{Dynamical processes and time evolution of distributions}
The distributions of $e_2$ and $q_3$ are consistent with the DI+DI formation scenario but there is also a need to account for the additional dynamical processes that exist in CHTs. The shift in $q_3$ needs strong mechanisms for mass loss which cannot be explained by stellar evolution only. \cite{Moharana_ST3n2} proposed mass-loss by tidal dissipation which would also explain the abundance of planar CHTs. But there is also a possibility that the different $q_3$ are a result of dynamical interactions with the circumbinary accretion disk (CBAD) or complicated fragmentation and accretion in the CBAD \citep{4EDV_Borko}. However, we lack a large sample of robust mutual inclination and age estimates to test these scenarios. \\

This deficiency will surely be improved with the upcoming large-scale missions that will study multiple stars. Further, our calculations are approximate and do not precisely estimate the $q_3$. This calls for detailed analyses of more CHTs and a better study of parameter distribution.

\acknowledgements
This work is funded by the Polish National Science Centre (NCN) through grant 2021/41/N/ST9/02746. A.M. acknowledges support from the UK Science and Technology Facilities Council (STFC) under grant number ST/Y002563/1.  K.G.H. and G.P. acknowledge support from NCN grant 2023/49/B/ST9/01671.

\bibliography{E12_Moharana}

\end{document}